\documentclass[aps,twocolumn,nofootinbib,longbibliography,prd]{revtex4-1}
\usepackage[babel]{csquotes}
\usepackage{graphicx}
\usepackage{amsmath,amssymb}
\usepackage[colorlinks,citecolor=blue,linkcolor=blue,urlcolor=blue]{hyperref}
\usepackage{mathrsfs}
\usepackage{enumerate}
\def\be{\begin{equation}}
\def\ee{\end{equation}}

\begin{document}

\title{Preparation in Bohmian Mechanics}

\author{Carlo Rovelli}
\affiliation{Aix-Marseille University, Universit\'e de Toulon, CPT-CNRS, F-13288 Marseille, France.}
\affiliation{Perimeter Institute, 31 Caroline Street N, Waterloo ON, N2L2Y5, Canada} 
\affiliation{The Rotman Institute of Philosophy, 1151 Richmond St.~N London  N6A5B7, Canada}

\begin{abstract}

\noindent  
According to Bohmian mechanics, we see the particle, not the pilot wave. But to make predictions we need to know the wave. How do we learn about the wave to make predictions, if we only see the particle?  I show that the puzzle can be solved, but only thanks to decoherence. 

\end{abstract}

\date{\small\today}

\maketitle

\subsection*{The problem}

In Bohmiam mechanics \cite{Goldstein2017}, the state of a system of $N$ degrees of freedom (or ``particles") is given by a couple $(\Psi(x_n),X_n)$, namely a wave function $\Psi(x_n)$, or ``pilot wave", and the positions $X_n$ of the particles. Here $n=1,...,N$.  The dynamics is given by the two equations
\be
i\hbar\;\partial_t \Psi = H\Psi,  \ \ \ \   \partial_t X_n=\frac{\hbar}{m_n}\left.\frac{\partial_{x_n}\Psi}{|\Psi|}\right|_{x_n=X_n}, 
\ee
where $H$ is the Hamiltonian operator and $m_n$ the mass of the $n$th particle.  The first is the Schr\"odinger equation, the second is Bohm's guidance equation. 

A theorem states that if at some time we do not know the position of the particles but we know their probability distribution and this satisfies $\rho({X_n})=|\Psi(X_n)|^2$, then the dynamics conserves this equation.   The value of this equation ``initially" is either assumed or said to be attained via a suitable equilibration process \cite{Valentini2010}, depending on the versions of the Bohmiam theory considered. In either case, if we know the wave function, we can give probabilistic predictions about where the particles will go.

According to the theory, what we see, are the actual positions $X_n$ of the variables (``particles").   Hence, in a sense, the particle positions are ``manifest" variables and the wave is ``hidden".   In a different sense, the particle positions are called the ``hidden variables", because they are variables added to conventional wave mechanics to make it (non local and) deterministic.

There is one aspect of this theory that I have never understood.  In order to make predictions, we have to know $\Psi$.   But how can we know $\Psi$, if what we see is only the particle positions $X_n$?    The theory gives us a probability distribution for the particle position, given the wave; but it does not give us a recipe for guessing the wave, given the particle positions that we see. (On the problem, see \cite{Bohm1984,Bohm1987}.)

In textbook quantum mechanics, the state is determined by a \emph{preparation}, which is essentially a measurement.  A measurement, in conventional quantum theory, \emph{collapses} the wave function on the eigenstate of the observed eigenvalue.   Hence we know the state because we made an observation, and we the theory instructs us to \emph{assume} that the state is then the corresponding eigenstate.  If we see spin up, we then say the state is $|+\rangle$.   

But nothing of the sort is directly true in Bohmiam mechanics: if we see the particle in the position $X$, the wave can still be virtually anything.   Hence we cannot infer much about the wave, from observing particles.   Hence we cannot measure the wave.  But then how can we make predictions, given that the wave is needed for predictions? 

Here, I give an answer to this question.   Maybe this answer is known, but I have not found it in the literature.  The answer is more subtle than what I expected, requires thinking about decoherence and does not require any hypothesis about special states in the past.

\subsection*{An insufficient solution}

In Bohmian literature, one reads that after a standard measurement the state of a System (with variable $s$) and  Apparatus (with pointer variable $a$) has the form 
\be
\Psi(s,a)=\sum_n c_v \ \psi_n(s)  \phi_n(a),
\label{meas}
\ee
where $n$ labels the eigenvalues of measured quantity, $\psi_n$ are the corresponding eigenstates,  and $\phi_n$ are states of the pointer variable that are peaked on well disjoint regions $R_n$ of its configuration space.  If the Bohmian pointer variable $a=A$ is in one of these regions, then the \emph{effective} state of the system $\psi(s)=\Psi(s,A)$ is in fact in the eigenstate of the measured quantity. 

All this is fine, of course, but  steps are missing.   The key one is that 
system and apparatus do \emph{not} end up in a state of the form \eqref{meas} if they are in a generic state before the measurement interaction.  This follows immediately from the fact the evolution during the interaction is unitary, hence a generic state ends up in a generic state, while the states \eqref{meas} are not generic.   To end up in the form \eqref{meas}, the combined system (system+apparatus) must be in a tensor state, namely uncorrelated, before the interaction.  But how can we know that it is so, if not by measurement?  So, to prepare a state, we already need to know a state. This appears to open a regress to infinity.   

Furthermore, even if we get to know the effective state, how do we know that other components of the universal wave function don't come-in, and affect the evolution governed by effective states? Here I try to fill these gaps.

\subsection*{The key preliminary corollary}

Consider a single particle.  The wave is $\Psi(x)$ and the particle position is $X$. Imagine that at some time the wave has two components disjoint in configuration space. That is $\Psi(x)=\Psi_1(x)+\Psi_2(x)$ where the support of $\Psi_1$ is a compact region $R_1$ and the support of $\Psi_2$ is a compact region $R_2$ that does not overlap with $R_1$. Then by the probability assumption the particle must be either in  $R_1$ or in $R_2$.  Say it is in $R_1$. Then, as long as $\Psi_1$ and $\Psi_2$ do not interfere, we can replace $\Psi$ with $\Psi_1$ in the guidance equation for $X$.

This is immediate, because the guidance equation depends only on the value of $\Psi$ in the neighbourhood of $x=X$ and $\Psi_2$ vanishes there.  Since the Schr\"odinger equation is linear, we can evolve $\Psi_1$ with this equation and disregard $\Psi_2$ entirely.  

All this stops to hold, of course, if there is interference, namely if at some time $R_1$ and $R_2$ get to overlap. 

Next consider the same single particle, but now interacting with an environment of many degrees of freedom $e_n$. The wave is $\Psi(x,e_n)$ and the positions are $(X,E_n)$. 
Imagine, as before, that at some time $\Psi$ has two components $\Psi_1$ and $\Psi_2$ such that the supports of $\Psi_1(x,E_n)$ and $\Psi_2(x,E_n)$ are disjoint in the space of the variable $x$.    Then, as before, we can disregard $\Psi_2$ for the dynamics of the particle, as long as there is no interference. 
But let's furthermore, imagine that the interaction between the particle and the environment is non negligible (in a precise sense specified below).  Then, interference is suppressed.  This is because the probability distribution for the particle position is given by
\begin{eqnarray}
\rho(x)&=&|\Psi(x,E_n)|^2=|\Psi_1(x,E_n)+\Psi_2(x,E_n)|^2\\ 	\nonumber
&=&|\Psi_1(x,E_n)|^2+ |\Psi_2(x,E_n)|^2 \\ 	\nonumber
&& +Re[\overline{\Psi_1(x,E_n)}\Psi_2(x,E_n)]. 
\end{eqnarray}
Interference is given by the last term, but if the coupling with the environment is sufficient, the two states $\Psi_1$ and $\Psi_2$ affect the environment differently and therefore $\Psi_1(x,e_n)$ and $\Psi_2(x,e_n)$ are going to have different support in the $e_n$ variables. But then 
there is no value of $E_n$ that can prevent this term from vanishing.  This is essentially the Bohmian version of decoherence. 

The consequence is that, once more, $\Psi_2$ can be discarded, as far the motion of $X$ is concerned, but now without limitations: there is no risk of interferences bringing back $\Psi_2$ into the dynamics of $X$.

\subsection*{The effective loss of quantum entanglement}

Consider now two systems, with position variables $S$ and $A$, and state $\psi(s,a)$, in an entangled state. That is
\be
\Psi(s,a)\ne \Psi(s) \Phi(a).
\ee
For simplicity, consider an entangled state of the form 
\be
\Psi(s,a)= \Psi_1(s,a)+\Psi_2(s,a)=  \Psi_1(s) \Phi_1(a)+\Psi_2(s) \Phi_2(a). 
\ee
Immagine that the system $A$ interacts sufficiently strongly with an environment.   Say $\Phi_1(a)$ and $\Phi_2(a)$ have disjoint support, as before,   and the variable $A$ is in the support of $\Phi_1(a)$. Then, for the argument above, the state $\Psi_2$ can be discarded, for the future evolution of the variable $A$ and $S$. 

This means that the interaction of a system $A$ with the environment has the effect to cancel any interference effect due to any entanglement between $A$ and another system $S$.  Once again, this can be recognised as a typical effect of decoherence.    Hence if $A$ interacts sufficiently with an environment, we can always effectively assume that its state is not entangled with another system $S$.

Armed with this, we can tackle the preparation of a state in Bohmian mechanics.  

\subsection*{Preparation}

Consider a variable $S$ (System) and a variable $A$ (apparatus, or pointer variable) interacting with an environment. Say the interaction hamiltonian between $S$ and $A$ depends on a function $O$ on the phase space of the coupled system, and is turned on at some time $t$.   Assume that before this time, $A$ has sufficiently interacted with the environment.  Then, because of the argument given above, we can assume that at time $t$ the state is a tensor state.   More precisely, as the variable $A$ has sufficiently interacted with the environment, the system and the apparatus can be effectively considered uncorrelated.  The empty components of the universal wave are there, but do not disturb.  This is the missing ingredient in previous analysis  of Bohmian measurement that I found. 

Let $\psi(s,a)=\Psi(s,a,E_n)$ be the effective state of System+Apparatus. Its evolution is effectively governed by a Schr\"odinger equation.  The standard Schr\"odinger evolution of an initial uncorrelated state $\psi(s,a)=\psi_A(a)\psi_S(s)$ by an interaction hamiltonian that depends on an observable $O$ drives the state to the well known form
\be
\psi(s,a)=\sum_n c_v \ \psi_n(s)  \phi_n(a),
\ee
as above, where $\phi_n$ are states of the pointer variable that are peaked on well disjoint regions $R_n$ of the configuration space of $a$.  In Bohm theory, the pointer $A$ is going to be in one of these, say the $n$th one.  By the results of the previous section, because of the fact that $A$ interacts with the environment, any interference between the terms of the last equation is suppressed and therefore the future evolution of the variable $S$ is going to be guided by the single eigenstate $\psi_n$ that correspond to the non empty branch of the apparatus wave.    This is preparation of a state in Bohmian mechanics.  

\subsection*{Conclusion}

Contrary to the expectation I had when I started looking into this, it works.  The process corresponding to decoherence plays a major role.  The trick is that it is true that knowledge of the $X_n$ gives us very little information about the $\Psi$, but this can be circumvented by having decoherence rendering most components of $\Psi$ irrelevant and arranging the dynamics so that the relevant part of the effective state $\psi$ is a finite sum of terms we know, so that the pointer variable being in one region restricts the effective wave function of the system to a known form. And this is of course exactly what experimenters do in a real laboratory experiment. 

To summarize: first, the environment-apparatus interaction renders the interference between states with different values of the pointer variable irrelevant.  Second, we can  write a \emph{generic} state of the combined (system+apparatus) as a sum of tensor terms peaked on different values of the pointer, and because of the previous point, we can forget the empty branches.  Hence we can safely assume that before the start of the interaction system and apparatus are uncorrelated. The measurement interaction correlates them. Finally, the value of the pointer value tells us precisely that the only relevant branch of the effective state of the system is the corresponding eigenstate. 

I embarked in this exploration thinking that Bohiam mechanics does not really work, but, as far as I can see, it works.  There is no need for this to make any special (``past low entropy") assumption (as I at some point suspected), but decoherence is necessary to make predictions.\\ 

\centerline{***}\vspace{2mm}

Many thanks to Wayne Myrvold, Davide Romano, and David Albert for comments, enlightening  discussions, and corrections to the first draft of this paper.

\bibliography{/Users/carlo/Documents/library/library}
\end{document}